\documentclass[pra,aps,showpacs,manuscript]{revtex4}
\usepackage{graphicx}
\begin{document}

\title{Manifestation of superfluidity in atom-number-imbalanced
two-component Bose-Einstein condensates}

\author{S. M. Al-Marzoug$^{1,2}$, B. B. Baizakov$^{3,4}$, U. Al Khawaja$^5$ and H. Bahlouli$^{1,2}$}
\affiliation{
$^{1}$ Interdisciplinary Research Center for
Intelligent Secure Systems, King Fahd University of Petroleum and Minerals, Dhahran 31261, Saudi Arabia,\\
$^{2}$ Physics Department, King Fahd University of Petroleum and Minerals, Dhahran 31261, Saudi Arabia, \\
$^{3}$ Physical-Technical Institute, Uzbek Academy of Sciences, 100084, Tashkent, Uzbekistan,\\
$^{4}$ Institute of Theoretical Physics, National University of Uzbekistan, Tashkent 100174, Uzbekistan,\\
$^{5}$ Department of Physics, School of Science, The University of Jordan, Amman, 11942, Jordan.
}
\date{\today}

\begin{abstract}
Superfluid and dissipative regimes in the dynamics of a
two-component quasi-one-dimensional Bose-Einstein condensate (BEC)
with unequal atom numbers in the two components have been explored. The
system supports localized waves of the symbiotic type owing to the
same-species repulsion and cross-species attraction. The minority
BEC component moves through the majority component and creates
excitations. To quantify the emerging excitations we introduce a
time dependent function called disturbance. Through numerical
simulations of the coupled Gross-Pitaevskii equations with periodic
boundary conditions, we have identified a critical velocity of the
localized wave, above which a transition from superfluid to
dissipative regime occurs, as evidenced by a sharp increase in the
disturbance function. The factors responsible for the discrepancy
between the actual critical velocity and the speed of sound,
expected from theoretical arguments, have been discussed.
\end{abstract}
\maketitle

\section{Introduction}

Superfluidity is the ability of a liquid to flow through tight tubes
or narrow slits without dissipation. The experimental discovery
\cite{kapitza1938,allen1938} and theoretical explanation
\cite{landau1941} of this macroscopic quantum phenomenon were among
the most important achievements of 20th-century physics. While
initial studies were restricted to liquid helium, superfluid
properties were recently observed and explored in a dilute gas of
bosons, called Bose-Einstein condensates (BEC)
\cite{pitaevskii-book,barenghi-book}. The advantage of ultracold
gases in studies of superfluidity over liquid helium is that they
are analytically tractable, weakly interacting systems, and free of
surface tension effects.

The phenomenon of superfluidity in quantum liquids and gases has
been studied in different contexts. Among them, the breakdown of
superfluidity when the flow velocity reaches some critical value is
particularly intriguing. The physics behind the suppression of
superfluidity and transition to a usual dissipative regime above the
critical velocity is linked to the generation of excitations in the
system such as phonons, rotons, and vortices when the liquid flows
at velocities exceeding some critical value $v_c$, which was predicted by
Landau \cite{landau1941,lifshitz-book} $v_c = {\rm
min}\left(\epsilon(p)/p\right)$, with $\epsilon(p)$ being the energy
of the excitation with momentum $p$. Measurements of the critical
velocity in BEC by laser stirring have confirmed that the
superfluidity indeed breaks down if the laser beam moves faster than
a certain finite speed \cite{raman1999,onofrio2000,engels2007}.
However, the measured values of the critical velocity ($\sim$ 1.6
mm/s) in the experiment with sodium BEC \cite{raman1999} appeared to
be smaller than the speed of sound ($\sim$ 6.2 mm/s), expected from
theoretical arguments. The diameter of the laser spot in the
experiment was macroscopic ($\sim 13 \, \mu {\rm m}$), exceeding the
intrinsic length scale of the condensate, so called healing length
$\xi = 1/\sqrt{8\pi a_s n} \sim 0.3 \, \mu {\rm m}$, where $a_s = 52
\, a_B$ - is the two-body $s$ - wave scattering length in units of
Bohr radius, $n = 1.5 \times 10^{14}$ cm$^{-3}$ - is the peak
density of the condensate. Under these conditions, vortex shedding
by the moving object \cite{kwon2015,kokubo2024}, rather than the
emission of phonons, was a more realistic phenomenon. Later
investigations have revealed several factors which could be
responsible for the observed discrepancy, such as the size and
strength of the probe object
\cite{desbuquois2012,kiehn2022,kwak2023}, the intrinsic nonlinearity
of BEC \cite{hakim1997,pavloff2002,leszczyszyn2009,abdullaev2012},
the system's dimensionality \cite{astrakharchik2004}, etc. All
arguments mentioned indicate the need for further research on the
origin of the critical velocity leading to the breakdown of
superfluidity in quantum gases. This is necessary to gain a better
understanding of this macroscopic quantum phenomenon.

This work aims to explore the manifestation of superfluid properties
in the dynamics of quasi-1D binary quantum gases. In our setting,
the mixture condensate can support stable localized waves of the
symbiotic type \cite{adhikari2005,perezgarcia2005} due to
same-species repulsion and cross-species attraction. We consider a
significant disparity in atom numbers between the components and a
toroidal trap potential for the binary condensate. Under these
conditions, the smaller localized component appears immersed in the
larger extended component, distributed over the whole integration
domain. The minority component is propelled by applying a potential
that only affects this component. The majority component is
subjected to a uniform (flat) potential, while the minority
component experiences a harmonic potential. This causes the minority
component to oscillate, similar to a pendulum, with zero velocity at
the turning points and maximum velocity at the center. By adjusting
the amplitude of these oscillations, we can observe different
scenarios where the majority component is either in a superfluid
state or not. In cases where the majority component is not in a
superfluid state, a sound wave is generated, and the energy of the
oscillating ``pendulum" is transformed into sound. In fact,
suppression of superfluidity and transition to a dissipative regime
occurs when the velocity of the probe object (localized smaller
component) reaches a critical value, leading to the emergence of
density waves superimposed on the uniform background of the larger
component. We introduce a measure called \textit{disturbance} to
quantify the emerging excitations in the majority component, whose
superfluidity is being tested.

\section{Model and governing equations}

We use the following coupled dimensionless Gross-Pitaevskii
equations (GPE) to describe the dynamics of a two-component BEC in a
quasi-1D trap potential
\begin{equation}\label{gpe}
i \frac{\partial \psi_i}{\partial t}+\frac{1}{2}\frac{\partial^2
\psi_i}{\partial x^2} + (g_i |\psi_i|^2 + g_{ij}|\psi_j|^2)  \psi_i
= V_i \psi_i, \qquad i,j = 1,2, \qquad i \not= j,
\end{equation}
where $\psi_i(x,t)$ are the mean field wave functions of the
condensate components, $g_i, g_{ij}$ are the same-species and
cross-species atomic interactions coefficients respectively, $V_i$
is external potential for the $i$-th component. Specifically, we use
a uniform potential for the majority component and a harmonic trap
for the minority component. Such a component-selective potential was
employed in studies of two-vortex collisions in particle-imbalanced
heteronuclear BEC mixtures \cite{richaud2023}. The dimensionless
quantities in these equations are scaled using the frequency of the
radial confinement $\omega_{\bot}$, atomic mass $m$ and radial
harmonic oscillator length $l_{\bot} = \sqrt{\hbar/m\omega_{\bot}}$
as follows: time $t \rightarrow t \omega_{\bot}$, space $x
\rightarrow x/l_{\bot}$, wave function $\psi_i \rightarrow
\sqrt{2|a_s|}\,\psi_i$, with $a_s$ being the atomic $s$-wave
scattering length. When considering imbalanced settings with
different number of atoms in the components $N_{1,2}=\int
|\psi_{1,2}(x)|^2 dx$, we shall denote $\psi_1(x,t)$ and
$\psi_2(x,t)$ as minority and majority components, respectively. The
system of coupled GPE (\ref{gpe}) is suitable for the description of
homonuclear BEC mixtures, such as two internal states of $^{87}$Rb
\cite{egorov2013}  or $^{39}$K \cite{semeghini2018}. In these
mixtures, all the atoms belong to the same element but can occupy
two different internal spin states. We use the periodic boundary
conditions for the GPE (\ref{gpe}) implying the condensate is held
in a toroidal trap potential similar to that reported in
\cite{ryu2007,ramanathan2011}.

\section{Numerical simulations}

\begin{figure}[htb]
\centerline{\qquad a) \hspace{8cm} b)}
\centerline{
\includegraphics[width=8cm,height=6cm,clip]{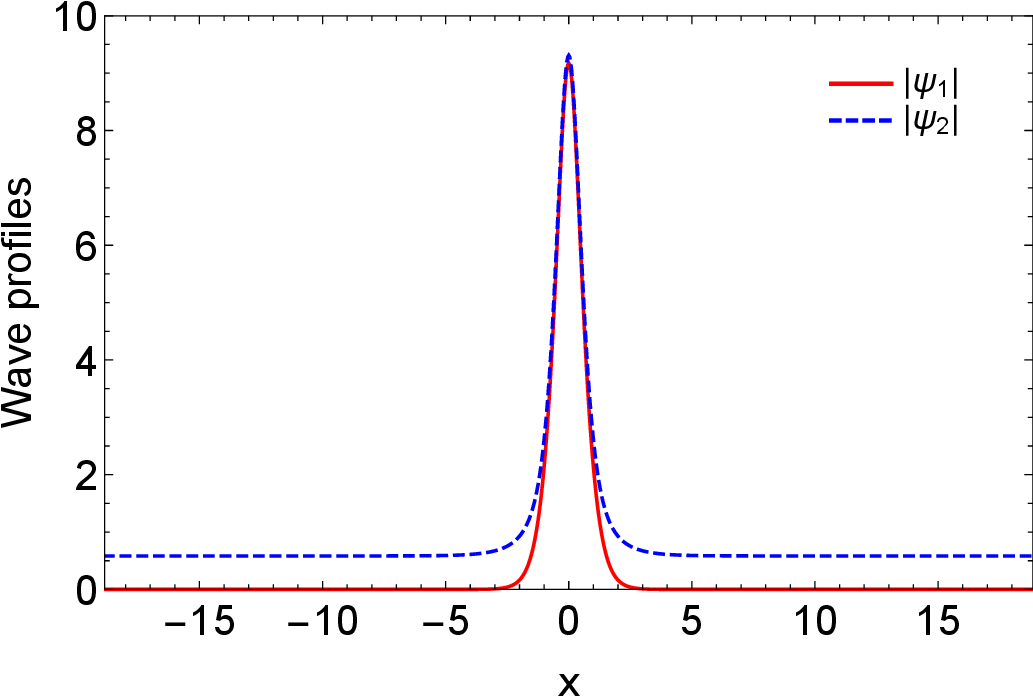}\qquad
\includegraphics[width=8cm,height=6cm,clip]{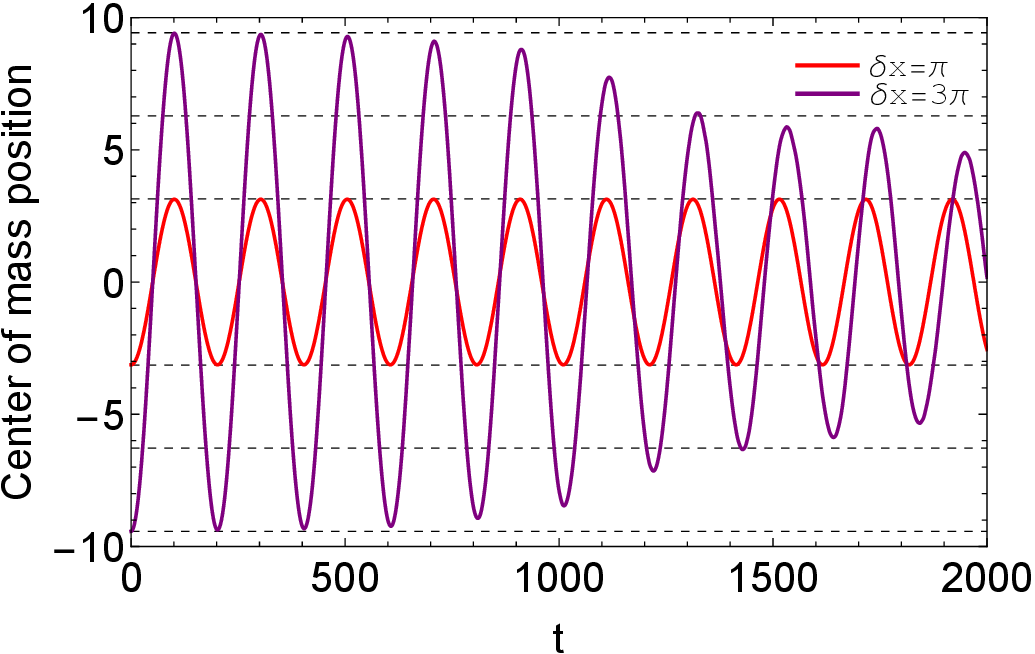}}
\caption{a) The initial state of the binary condensate with
same-species repulsion ($g_{1}=g_{2}=-1$) and cross-species
attraction ($g_{12}=g_{21}=1.05$) obtained by numerical solution of
the GPE (\ref{gpe}) with periodic boundary conditions. The smaller
component $\psi_1$ forms a localized wave while the larger component
$\psi_2$ spreads over the whole integration domain. b) The
center-of-mass position of the minority component $\psi_1$,
initially shifted by $\delta x$ from the minimum of the harmonic
trap $V_1(x)$ and released, performs oscillations near the origin.
At moderate displacement ($\delta x = \pi$) it shows fully
conservative/superfluid dynamics (red), while strong displacement
($\delta x = 3\pi$) leads to damped oscillations (purple) evidencing
the absence of superfluidity. Parameter values: $N_1=80$, $N_2=100$,
$V_1(x) = 0.001\, x^2, V_2(x) = 0$.} \label{fig1}
\end{figure}

Numerical simulations start with the creation of a localized state
in the particle imbalanced ($N_1 \not= N_2$) binary condensate with
repulsive intra-component ($g_1 < 0, g_2 <0$) and attractive
inter-component ($g_{12} > 0$) interactions, as shown in Fig.
\ref{fig1}a. The density of the majority component $|\psi_2|^2$
beyond the space occupied by the localized wave $|\psi_1|^2$ is
relatively small and uniform. Here, we are interested in the
superfluid and dissipative behaviors exhibited by the majority
component.

In the next step, we displace the minority component by $\delta x$
from its initial position at $x=0$ and let it oscillate in the
harmonic potential $V_1(x) = \beta x^2$, as illustrated in
Fig.~\ref{fig1}b. The center-of-mass position of the minority
component was evaluated according to its definition
\begin{equation}\label{eta}
\eta(t) =\frac{1}{N_1}\int_{-L/2}^{L/2} x |\psi_1(x,t)|^2 dx,
\end{equation}
where $\psi_1(x,t)$ is the solution of GPE (\ref{gpe}) at the given
time instance. By analyzing the oscillation dynamics of the minority
component for different initial displacements we have revealed a
critical value $\delta x_{cr} \sim 2.5\, \pi$ below which the
dynamics are fully conservative, while larger displacements
demonstrate damped oscillations, as shown in Fig. \ref{fig1}b
respectively for $\delta x = \pi$, and $\delta x = 3\pi$. These
numerical experiments emulate the shuttle motion of a probe object
or laser beam in a superfluid. It is natural to expect that at
sufficiently large oscillation amplitudes, the probe object acquires
a velocity greater than superfluid critical velocity ($v>v_c$) near
the minimum of the harmonic trap at $x=0$, producing excitations in
the condensate. Table 1 presents the velocity of the localized
minority component when it passes through the minimum of the
harmonic trap for different initial displacements
\begin{table}[h]
\begin{tabular}{|c|c|c|c|c|c|}
  \hline
  Initial shift, $\delta x$ & $\pi$ & $1.5 \, \pi$ & $2 \, \pi$ & $2.5 \, \pi$ & $3 \, \pi$ \\
  \hline
  Max. velocity, $v$ & 0.10 & 0.15 & 0.19 & 0.24 & 0.29 \\
  \hline
\end{tabular}
\caption{The maximal velocity $v$ acquired by the minority component
in the harmonic trap for different initial shifts $\delta x$,
according to numerical solution of the GPE (\ref{gpe}).}
\end{table}

\begin{figure}[htb]
\centerline{\qquad a) \hspace{8cm} b)} \centerline{
\includegraphics[width=8cm,height=6cm,clip]{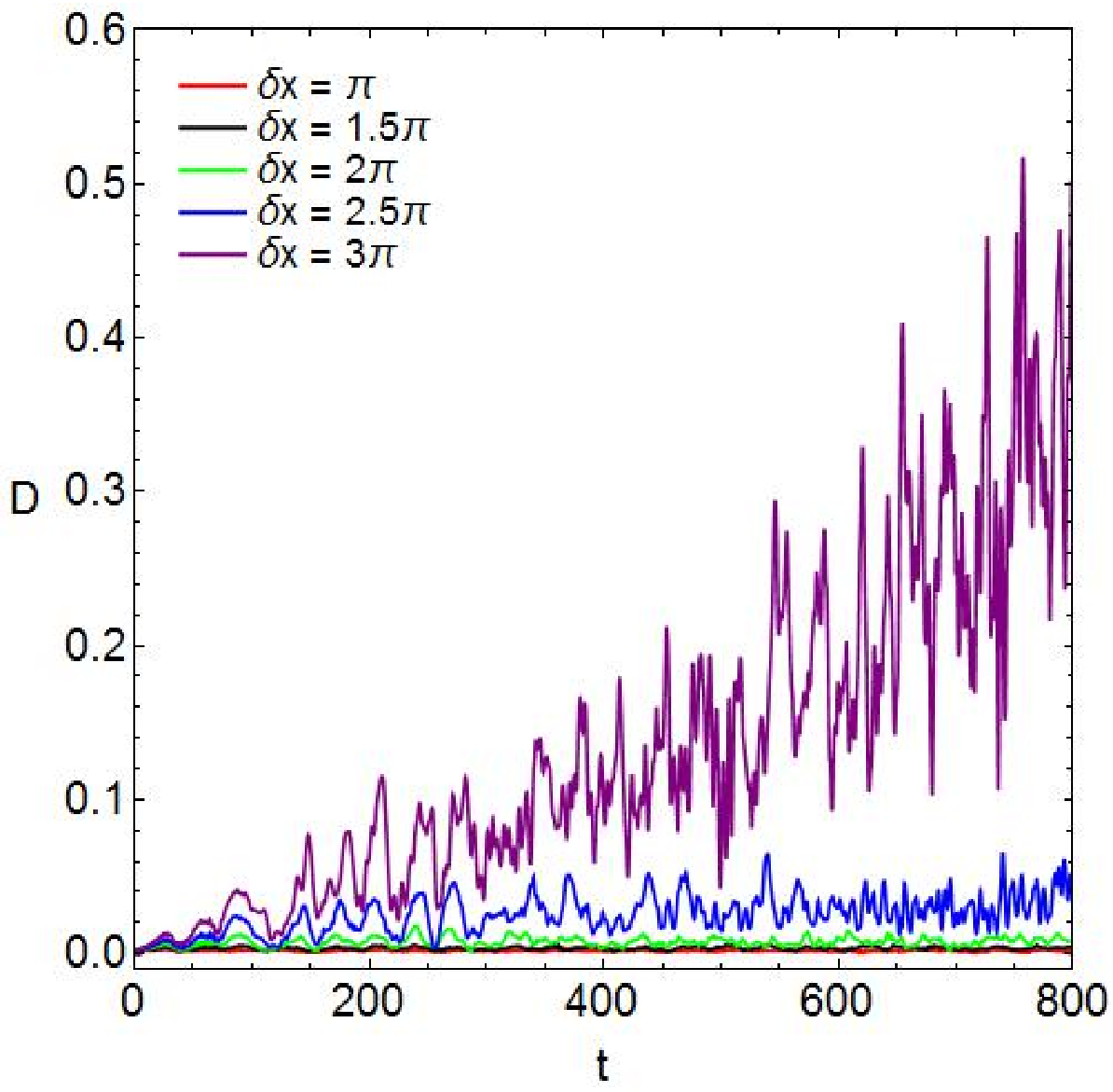}\qquad
\includegraphics[width=8cm,height=6cm,clip]{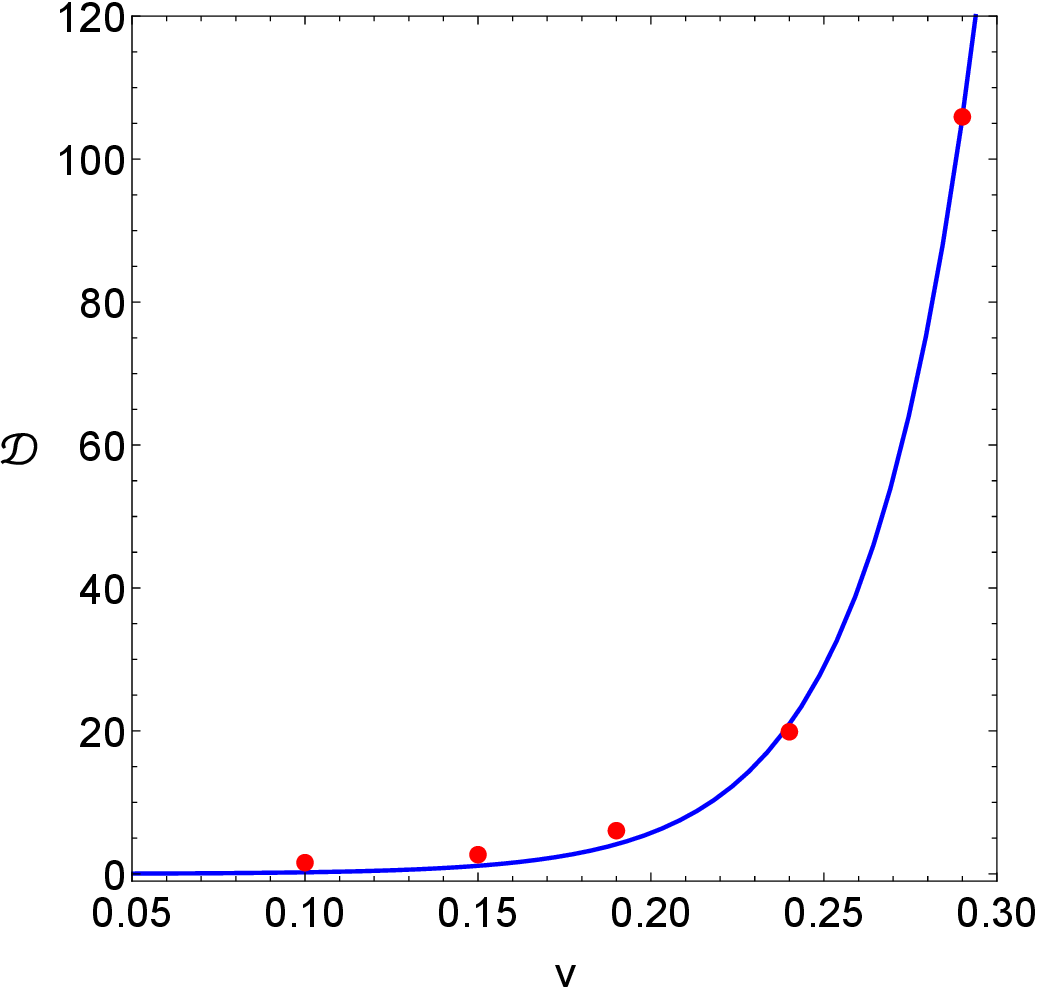}}
\caption{a) Growth of excitations in the majority component
($\psi_2$) during oscillations of the minority component ($\psi_1$)
in the harmonic trap for different initial displacements $\delta x$
according to Eq. (\ref{D}). A sharp transition from superfluid to
dissipative regime is observed near the critical shift $\delta x_{c}
\simeq 2.5\, \pi$. b) The integral disturbance ${\cal D} = \int
\limits_0^{t_f} D(t) dt$ as a function of maximal velocity $v$
attained by the minority component during oscillations in the
harmonic trap ($t_f = 800$). Data for the velocity (red points) are
taken from Table 1, while the blue curve represents the exponential
model ${\cal D}(v) = \alpha e^{\gamma \, v }$ with $\alpha = 0.009,
\gamma = 32.4$. } \label{fig2}
\end{figure}

We introduce a time-dependent function called \textit{disturbance}
to quantify the generation and growth of excitations in the
background condensate of the majority component
\cite{paris-mandoki2017}
\begin{equation}\label{D}
D(t) = \int_C (|\psi_2(x,t)|^2 - |\psi_2(x,0)|^2)^2\, dx,
\end{equation}
where the integration is performed over the spatial domain with
vanishing amplitude of the smaller component $C \in
\left[|\psi_1(x,t)| \simeq 0 \right]$. The latter condition ensures
we consider only the excitations on top of the uniform part of the
repulsive condensate $\psi_2$. Figure \ref{fig2} shows the growth
rate of excitations for various initial displacements of the
minority component in relation to the minimum of the harmonic trap.

As can be seen from Fig. \ref{fig2} a, the movement of the minority
component doesn't produce excitations in the majority component,
thus the disturbance function Eq. (\ref{D}) does not grow with time,
if its initial position is not sufficiently far from the origin
($\delta x < 2.5 \, \pi$). In contrast, at a greater initial shift
($\delta x = 3 \, \pi$), the disturbance function rapidly grows with
time, evidencing the vigorous generation of excitations. These
excitations in the majority component appear as density modulations
(sound waves) on the initially uniform domain of $\psi_2$. It is
reasonable to suggest that the generation rate of quasiparticles or
density waves starts to grow exponentially once the velocity of the
probe object exceeds the critical value. Figure \ref{fig2}b shows
the behavior of the integral disturbance ${\cal D}$ for different
maximal velocities of the minority component moving through the
majority component. As can be seen from this figure, numerically
obtained data (red points) nicely fit with the exponential model
${\cal D}(v) = \alpha e^{\gamma \, v }$.

\begin{figure}[htb]
\centerline{\qquad a) \hspace{8cm} b)}
\centerline{
\includegraphics[width=8cm,height=6cm,clip]{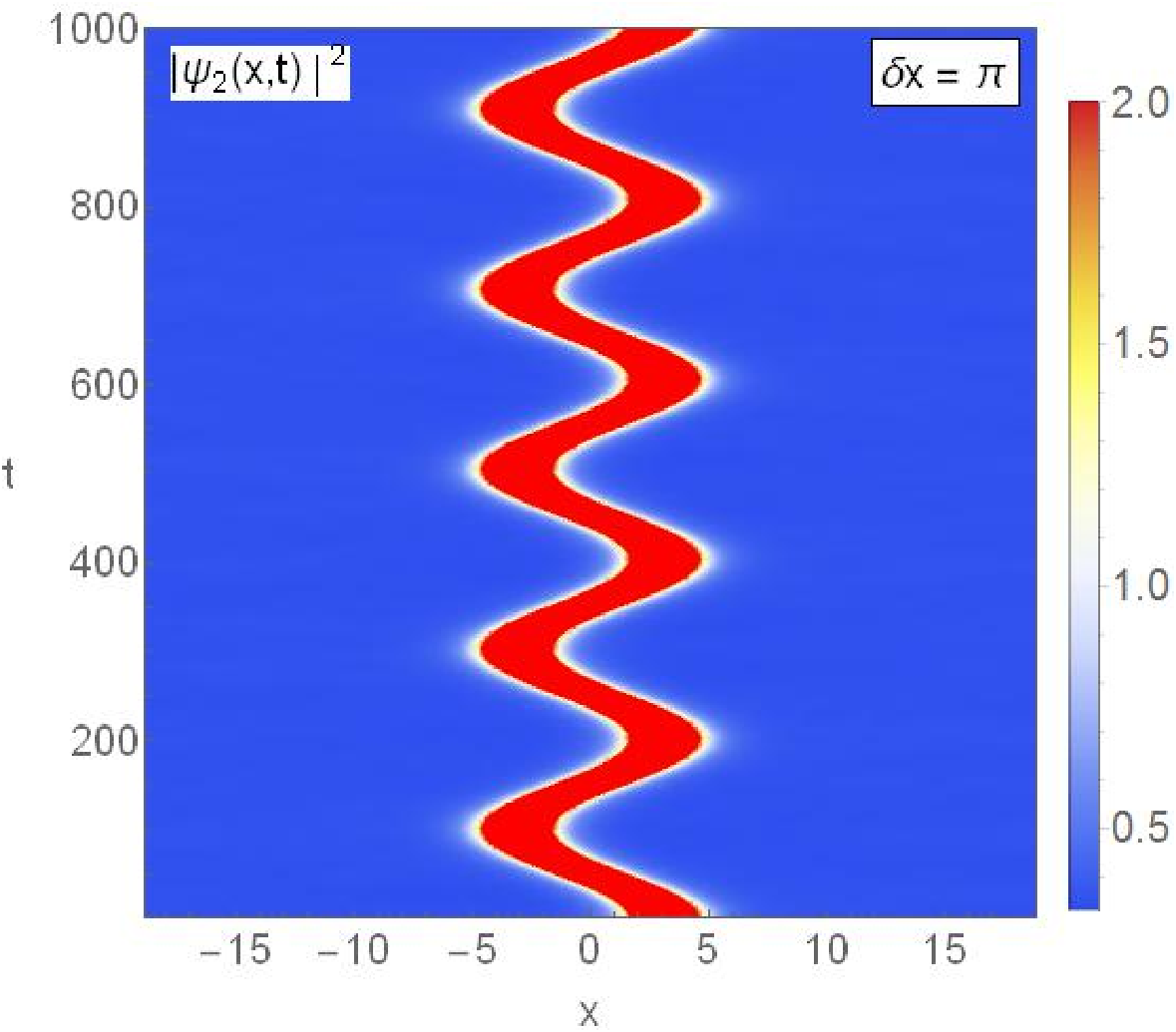}\qquad
\includegraphics[width=8cm,height=6cm,clip]{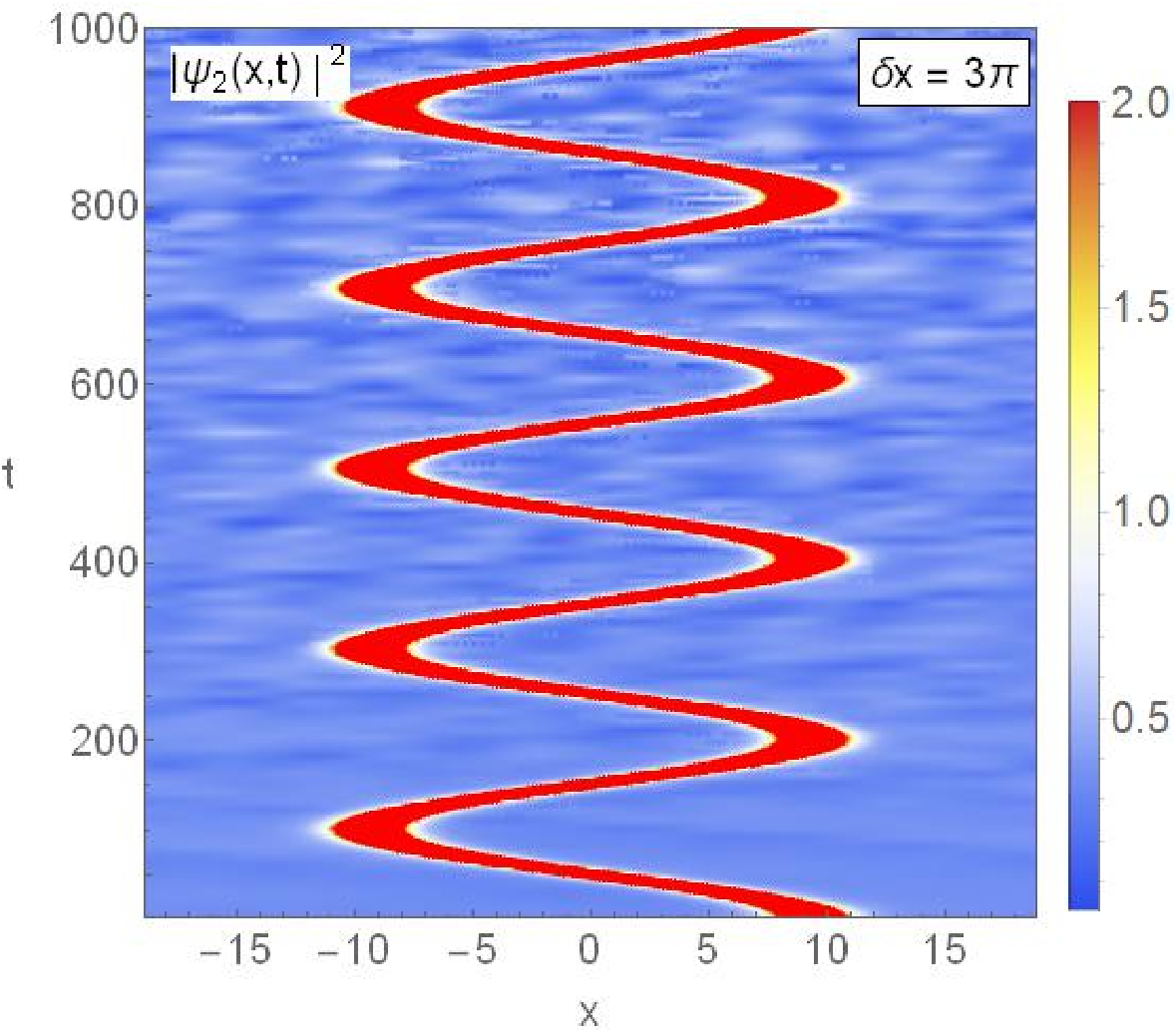}}
\centerline{\qquad c) \hspace{8cm} d)} \centerline{
\includegraphics[width=7cm,height=6cm,clip]{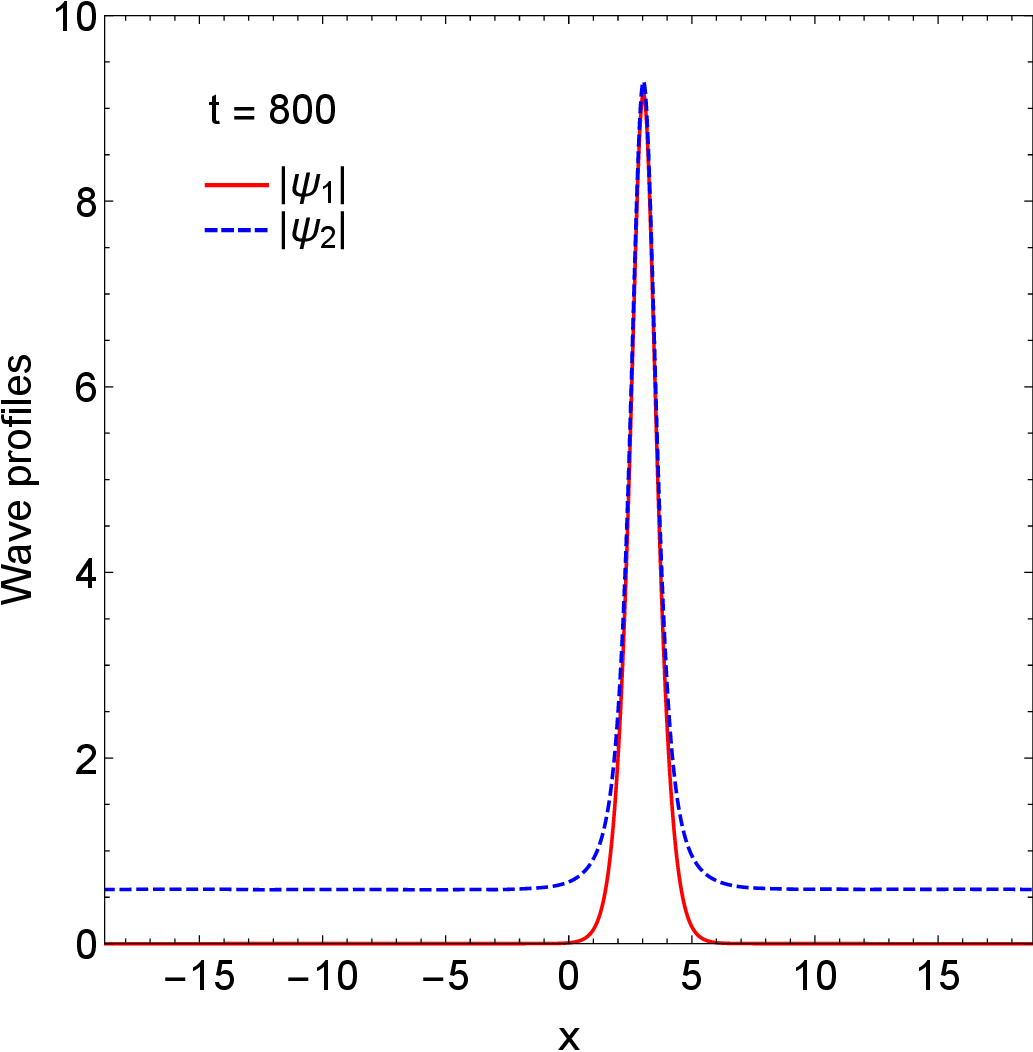}\qquad \qquad
\includegraphics[width=7cm,height=6cm,clip]{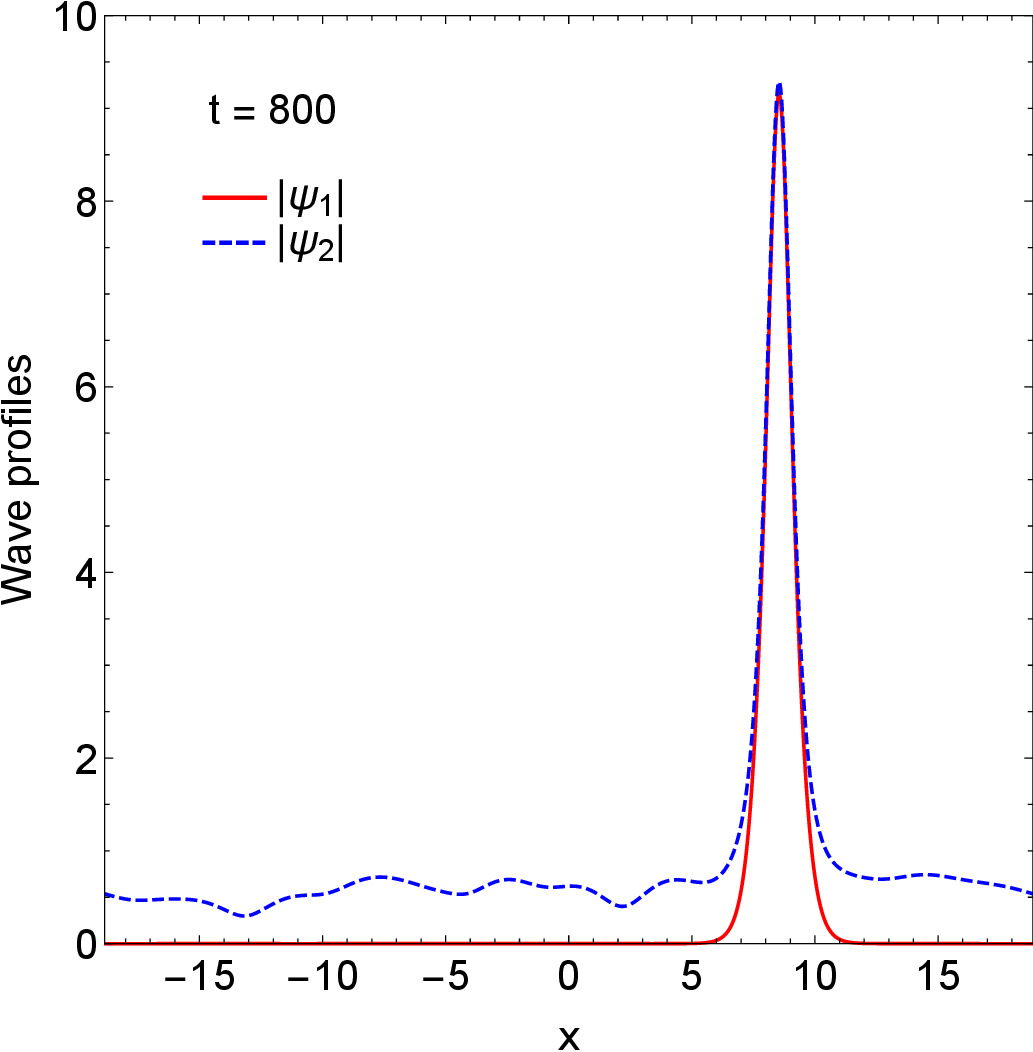}}
\caption{The density modulations on the background condensate
$\psi_2(x,t)$ do not appear when the minority component performs
small amplitude oscillations ($\delta x = \pi$), attaining
subcritical velocity $v = 0.1$ near the bottom of the parabolic trap
(a, c). At large amplitude oscillations ($\delta x = 3 \pi$) the
probe object attains supercritical velocity $v = 0.29$, thus the
density modulations strongly amplify (b, d). To highlight the
density waves in the condensate, only the low-intensity part
$|\psi_2(x,t)|^2 < 2$ has been shown (a, b).} \label{fig3}
\end{figure}

Different dynamical regimes can be explored by altering either the
initial position or the initial velocity of the probe object in the
harmonic trap. The simulation results presented in Fig. \ref{fig3}
demonstrate the transition from the superfluid regime to the
dissipative regime when the initial displacement of the minority
component exceeds the critical value. The generation of excitations
occurs during time intervals when the localized minority component
repeatedly passes the minimum point of the harmonic potential with
supercritical velocity. It is evident that the density modulations
on the background condensate $|\psi_2(x,t)|^2$ do not emerge when
the minority component undergoes small-amplitude oscillations and
reaches subcritical velocity near the bottom of the parabolic trap,
as shown in Fig. \ref{fig3}\,a,c. However, during large-amplitude
oscillations, the probe object reaches supercritical velocity,
causing the density modulations on $|\psi_2(x,t)|^2$ to strongly
amplify (see Fig. \ref{fig3}\,b,d).

\begin{figure}[htb]
\centerline{\qquad a) \hspace{8cm} b)}
\centerline{
\includegraphics[width=8cm,height=6cm,clip]{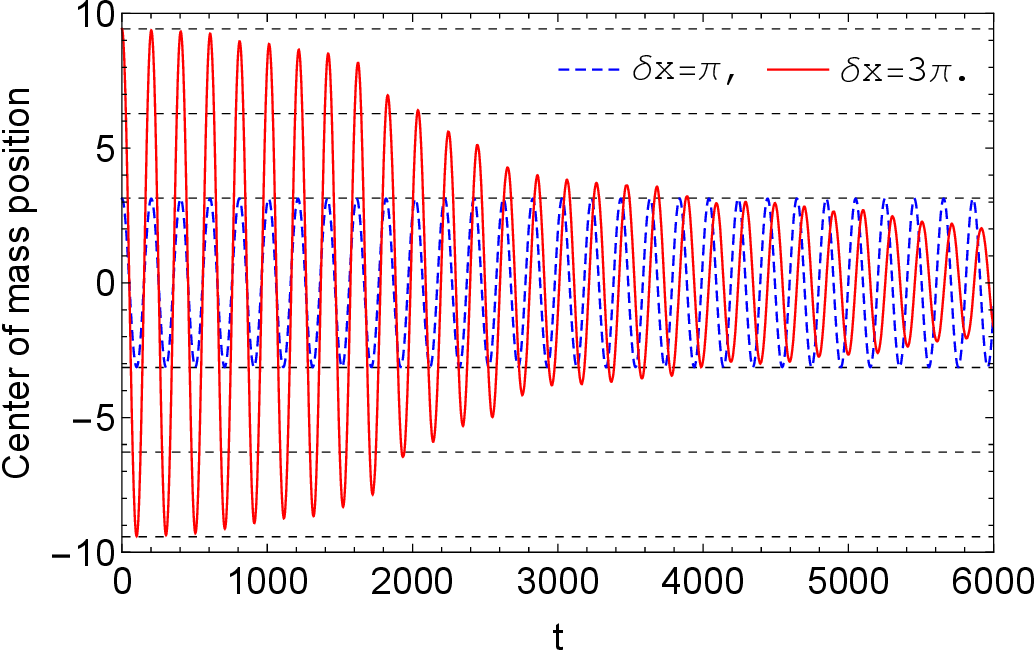}\quad
\includegraphics[width=8cm,height=6cm,clip]{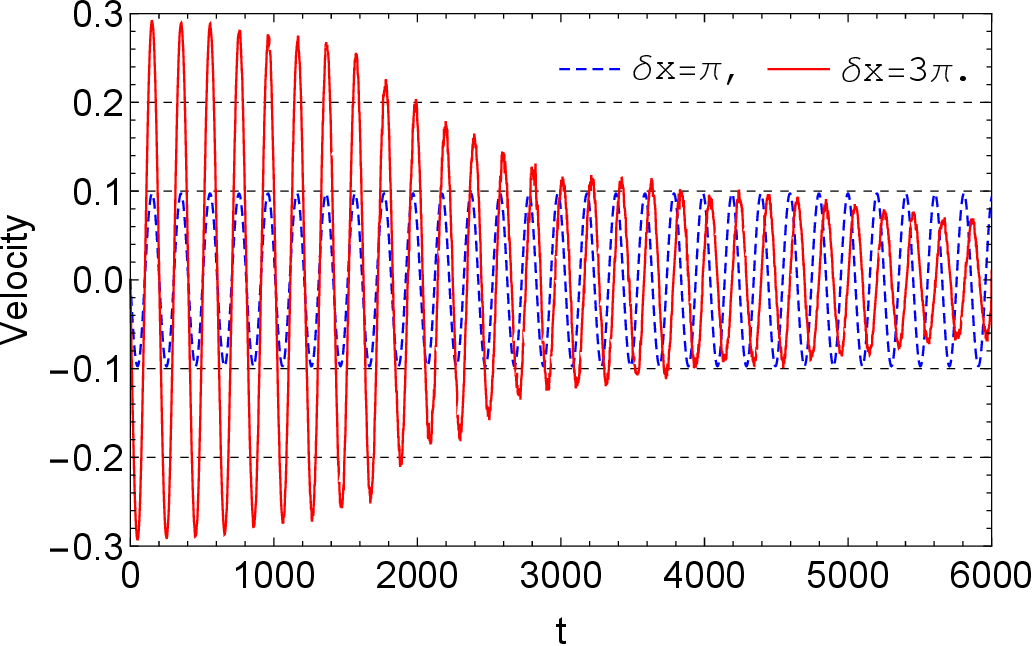}}
\caption{Oscillations of the center-of-mass position $\eta(t)$ (a)
and velocity $d \eta/dt$ (b) of the minority component, which was
initially shifted from the minimum of the harmonic trap, according
to Eq. (\ref{eta}). At a smaller displacement ($\delta x = \pi$),
the localized wave undergoes free oscillations with a constant
amplitude, indicating the presence of the superfluid regime (blue
dashed line). In comparison, a larger displacement ($\delta x =
3\pi$) leads to damped oscillations, indicating the onset of the
dissipative regime (red solid line). Later, the superfluid regime is
reestablished, with some reduction in energy of the minority
component due to the quasiparticles created in the majority
component.} \label{fig4}
\end{figure}

The existence of a definite superfluid critical velocity $v_c$ can
be demonstrated by examining the long-time evolution of the
oscillating minority component whose initial condition was in the
dissipative domain. This implies that it is far shifted from the
minimum of the harmonic trap and initially performs large amplitude
oscillations, thus attaining a supercritical velocity ($v>v_c$) at
the origin. It is reasonable to expect, that after gradually losing
its kinetic energy due to generation of excitations in the larger
component, the probe object slows down attaining the subcritical
velocity ($v<v_c$), and the superfluid regime is reestablished. To
illustrate this idea in Fig. \ref{fig4} we show the long-time
evolution of the center-of-mass position and velocity of the
minority component for two initial displacements, corresponding to
the superfluid and dissipative regimes. As can be seen from this
figure, at a smaller displacement ($\delta x = \pi$), the localized
wave undergoes free oscillations with a constant amplitude,
indicating the presence of the superfluid regime (blue dashed line).
On the other hand, a larger displacement ($\delta x = 3\pi$) results
in damped oscillations, suggesting the onset of the dissipative
regime (red solid line). Subsequently, after evolution time $t>3000$
the superfluid regime is reestablished, albeit with some reduction
in energy of the minority component due to the excitations created
in the majority BEC component.

Numerical simulations were performed using the methods of split-step
fast Fourier transform \cite{agrawal-book,numrecipes} and a
Runge-Kutta in interaction picture \cite{hult2007}. The integration
domain of length $L \in [-6\pi, 6\pi]$ has been employed to
accommodate 1024 Fourier modes with a corresponding space step
$\Delta x = 0.036$. The time step was $\Delta t = 0.0005$. The
periodic boundary conditions $\psi_i(-L/2,t)=\psi_i(L/2,t)$,
$i=1,2$, were adopted to emulate a toroidal trap configuration. The
stationary states of the mixture condensate described by GPE
(\ref{gpe}) were constructed using the Pitaevskii damping procedure
\cite{choi1998}. The stability of ground-state solutions was
confirmed by observing the long-term evolution of weakly perturbed
initial wave profiles. Different initial conditions for the
governing GPE were prepared by shifting the position of the minority
component in the harmonic trap relative to its minimum at $x=0$.

Now we will estimate the parameter values used for numerical
simulations in physical units and compare them with previously
reported experimental data. Let us consider a BEC of $^{87}$Rb atoms
prepared in two internal states $|F=1, m_F=-1\rangle$ and $|F=2,
m_F=1\rangle$. The $s$-wave scattering lengths of rubidium atoms in
these ground hyperfine states are almost equal $a_{1}\simeq a_{2}
\simeq 100 \, a_B$, while the inter-component coefficient $a_{12}$
can be independently tuned. The transfer of atoms from one state to
the other can be induced by coherent electromagnetic radiation until
the desired populations $N_1$, $N_2$ in corresponding states
$\psi_1(x,t)$ and $\psi_2(x,t)$ are achieved. The mixture condensate
is supposed to be held in a toroidal trap with a transverse
confinement frequency $\omega_{\bot} = 2 \pi \times 100$ Hz. Then
the units of time and space are given by $\tau = 1/\omega_{\bot}
\simeq 1.6$ ms, $a_{\bot}= \sqrt{\hbar/m \omega_{\bot}} \simeq 1\,
\mu$m, respectively. The length of the integration domain
(circumference of the toroidal trap) is equal to $L=12 \, \pi
a_{\bot} \simeq 41 \, \mu$m. The density of the uniformly
distributed condensate's majority component consisting of ${\cal
N}_1=10^4$ rubidium atoms, required for calculation of the sound
velocity and healing length is $n_1 = {\cal N}_1/L a_{\bot}^2 \simeq
2.1 \times 10^{20}$~m$^{-3}$. Using these data one can estimate the
speed of sound $c_s = (4 \pi \hbar^2 |a_s| n_1/m^2)^{1/2} \simeq
2.7$ mm/s and the healing length $\xi = 1/\sqrt{8\pi |a_s| n_1}
\simeq 0.19 \, \mu {\rm m}$.

It is important to note that the above presented values are
associated with uniformly distributed BEC components, taking place
for $g_{12}=0$. A localized state of symbiotic type emerges when
there is attraction between the components ($g_{12} = -1.05$),
leading to a decrease in the density of the majority component away
from the minority one. For a particular case shown in
Fig.~\ref{fig1}a the density is reduced by approximately three
times. Thus we obtain the corrected values $c_s \simeq 1.6$ mm/s and
$\xi \simeq 0.32 \, \mu {\rm m}$.

It is evident from Fig. \ref{fig2}b that the transition from a
superfluid to a dissipative regime occurs near the critical velocity
$\sim 0.1$, which in physical units corresponds to $v_c \simeq 0.07$
mm/s. Therefore the critical velocity leading to suppression of
superfluidity in BEC appears to be significantly smaller than the
sound velocity, their ratio being $v_c/c_s \simeq 0.04$. Similar
ratios were reported in the experiments with sodium
\cite{onofrio2000} $v_c/c_s \simeq (0.07 \pm 0.02)$ and rubidium
condensates \cite{raman1999} $v_c/c_s \simeq 0.25$.

The reasons why the theoretical prediction overestimates the actual
critical velocity were addressed in many previous works (see e.g.
\cite{kiehn2022}). Among them are the inhomogeneity of the medium,
the influence of nonlinearity, macroscopic size of the moving object
and dimensionality of the system. All quoted factors are inherent to
our model. In particular, the size of the localized wave (moving
minority component) calculated for the symmetric setting
\cite{ismailov2024} ($N = N_1 = N_2, \ g = g_1 = g_2$) $a_0 =
\sqrt{2\pi}/[(g+g_{12}) N] \simeq 1.2 \, \mu$m was greater than the
healing length $\xi \simeq 0.3 \, \mu {\rm m}$. Thus the probe
object in our model is macroscopic. For now, uncovering the physical
mechanisms behind the onset of the dissipative regime in a BEC
superfluid and defining the associated critical velocity is still a
challenge.

\section{Conclusions}

Through numerical simulations, we have studied the superfluid and
dissipative regimes in the dynamics of a binary Bose-Einstein
condensate where the localized minority component, acting as a
single entity (probe object), moves through the majority component
and hence creating excitations. We have designed a configuration in
which the minority component is surrounded by the majority one and
experiences regular back-and-forth movements under the action of a
harmonic trap. Its velocity varies from zero at classical turning
points to a maximum value at the lowest point of the harmonic trap.
When the velocity of motion exceeds some critical value, a strong
dissipation occurs, accompanied by a production of quasiparticles in
the majority component. After multiple oscillations, the minority
component experiences energy losses and slowing down, which
eventually cause the return to a non-dissipative mode. It was found
that the actual value of the critical velocity leading to the
suppression of superfluidity in BEC is smaller than the speed of
sound expected from theoretical arguments. The possible reasons for
the observed discrepancy were analyzed.

\section*{Acknowledgements}

The authors gratefully acknowledge the support provided by the
Interdisciplinary Research Center for Intelligent Secure Systems
(IRC-ISS) at King Fahd University of Petroleum and Minerals (KFUPM)
in funding this work through project No. INSS2302.

\end{document}